\documentclass[11pt, a4paper, nosuperscriptaddress, showpacs, titlepage,showkeys]{revtex4-1}
\usepackage{times}
\usepackage{amsmath}
\usepackage{amssymb}
\usepackage{graphicx}
\usepackage{subfigure}
\usepackage{booktabs}
\usepackage{rotating}
\usepackage{longtable}
\usepackage{bm}
\makeatletter

\newcommand{\Rmnum}[1]{\expandafter\@slowromancap\romannumeral #1@}
\makeatother

\begin{document}

\title{Semi-exact solutions of the Razavy potential}
\author{Qian Dong$^1$}
\author{F. A. Serrano$^{2}$}
\author{Guo-Hua Sun$^3$}
\author{Jian Jing$^4$}
\author{Shi-Hai Dong$^1$}
\email[Corresponding author:]{dongsh2@yahoo. com (S. H. Dong). }
\affiliation{$^1$ Laboratorio de Informaci\'{o}n Cu\'{a}ntica, CIDETEC, \\
Instituto Polit\'{e}cnico Nacional, UPALM, CDMX 07700, Mexico\\
$^2$ Escuela Superior de Ingenier\'{\i}a Mec\'{a}nica y El\'{e}ctrica UPC, Instituto Polit\'{e}cnico Nacional, Av. Santa Ana 1000, M\'{e}xico, D. F. 04430, Mexico\\
$^3$ Catedr\'{a}tica CONACYT, CIC, Instituto Polit\'{e}cnico Nacional, CDMX 07738, Mexico\\
$^4$Department of Physics and Electronic, School of
Science, Beijing University of Chemical Technology, Beijing 100029,
P. R. China}
\pacs{03. 67. Mn, 03. 65. Ud, 04. 70. Dy}


\begin{abstract}
In this work we study the quantum system with the symmetric Razavy potential and show how to find its exact solutions.
We find that the solutions are given by the confluent Heun functions. The eigenvalues have to be calculated numerically. 
The properties of the wave functions depending on $m$ are illustrated graphically for a given potential
parameter $\xi$. We find that the even and odd wave functions with definite parity are changed to odd and even wave functions when the potential parameter $m$ increases.
This arises from the fact that the parity, which is a defined symmetry for very small $m$, is completely violated for large $m$.
We also notice that the energy levels $\epsilon_{i}$ decrease with the increasing potential parameter $m$.

\end{abstract}
\maketitle

\section{Introduction}
It is well known that the exact solutions of quantum systems play an important role since
the early foundation of the quantum mechanics. Generally speaking, two typical examples are studied for the hydrogen atom and harmonic oscillator
in classical quantum mechanics textbooks [1, 2]. Up to now, there are a few main methods to solve the quantum soluble systems.
The first is called the functional analysis method. That is to say, one solves the second-order differential equation and obtains their solutions [3], which are
expressed by some well-known special functions. The second is called the algebraic method, which is realized by studying the Hamiltonian
of quantum system. This method is also related with supersymmetric quantum mechanics (SUSYQM)[4], further closely with the factorization method [5].
The third is called as the exact quantization rule method [6], from which we proposed proper
quantization rule [7], which shows more beauty and symmetry than
exact quantization rule. It should be recognized that almost all soluble potentials mentioned above belong to single well potentials. The double well potentials have not been
studied well due to their complications [8-17], in which many authors have been searching the solutions of the double-well potentials for a long history. This is because
the double well potentials could be used in the quantum theory of molecules to describe the motion of the particle in the presence of two centers of force, the heterostructures, Bose-Einstein condensates and superconducting circuits, etc.

Almost forty years ago, Razavy proposed a bistable potential [18]
\begin{equation}\label{pon}
V(x)=\frac{\hbar^2\beta^2}{2\mu}\left[\frac{1}{8}\xi^2\cosh(4\beta x)-(m+1)\xi\cosh(2\beta x)-\frac{1}{8}\xi^2\right],
\end{equation} which depends on three potential parameters $\beta$, $\xi$ and a positive integer $m$. In Fig. 1 we plot it as the function of the variables $x$ with various $m$, in which we take $\beta=1$ and $\xi=3$.
Choose atomic units $\hbar=\mu=1$ and also take $\mathcal{V}(x)=2V(x)$. Using series expansion around the origin, we have
\begin{equation}\label{}
\mathcal{V}(x)=(-m \xi -\xi )+x^2 \left(-2 m \xi +\xi ^2-2 \xi \right)+\frac{2}{3} x^4 \left(-m \xi +2 \xi
   ^2-\xi \right)+\frac{4}{45} x^6 \left(-m \xi +8 \xi ^2-\xi \right)+O\left(x^7\right),
\end{equation} which shows that $\mathcal{V}(x)$ is symmetric to variable $x$. We find that the minimum value of the potential $\mathcal{V}_{\rm min}(x)=-(m+1)^2-\xi ^2/4$ at two minimum values $x=\pm\frac{1}{2}\cosh ^{-1} [2 (m+1)/\xi]$. For a given value $\xi=3$, we find that the potential has a flat bottom for $n=0$, but for $n>1$ it takes the form of a double well. Razavy presented the so-called exact solutions by using the "polynomial method" [18].  After studying it carefully, we find that the solutions cannot be given exactly
due to the complicated three-term recurrence relation. The method presented there [18] is more like the Bethe Ansatz method as summarized in our recent book [19].
That is, the solutions cannot be expressed as one of special functions because of three-term recurrence relations.
In order to obtain some so-called exact solutions, the author has to take some constraints on the coefficients in the recurrence relations as shown in Ref. [18].
Inspired by recent study of the hyperbolic type potential well [20-28], in which we have found that
their solutions can be exactly expressed by the confluent Heun functions [23], in this work we attempt to study the solutions of the Razavy potential. We shall find that
the solutions can be written as the confluent Heun functions but their energy levels have to be calculated numerically since the energy term is involved within the parameter $\eta$
of the confluent Heun functions $H_{c}(\alpha, \beta, \gamma, \delta, \eta, z)$. This constraints us to use the traditional Bathe ansatz method
to get the energy levels. Even though the Heun functions have been studied well, its main topics are focused in the mathematical area. Only recent connections with the physical problems have been
discovered, in particular the quantum systems for those hyperbolic type potential have been studied [20-28].
The terminology "semi-exact" solutions used in Ref. [21] arises from the fact that the wave
functions can be obtained analytically, but the eigenvalues cannot be written out explicitly.

This paper is organized as follows. In Sect. 2, we present
the solutions of the Schr\"{o}dinger equation with the Razavy potential. It should be recognized that the Razavy potential is single or double well depends on the potential
parameter $m$. In Sect. 3 some fundamental properties of the solutions
are studied. The energy levels for different $m$ are calculated numerically. Some concluding remarks
are given in Sect. 4.

\section{Semi-exact solutions}
Let us consider the one-dimensional Schr\"{o}dinger equation,
\begin{equation}\label{sch}
-\frac{\hbar^2}{2\mu}\frac{d^2}{dx^2}\psi(x)+V(x)\psi(x)=E\psi(x).
\end{equation}

Substituting potential (\ref{pon}) into (\ref{sch}), we have
\begin{equation}\label{psi}
\frac{d^2}{dx^2}\psi(x)+\left\{\varepsilon-\left[\frac{1}{8}\xi^2\cosh(4x)-(m+1)\xi\cosh(2x)-\frac{1}{8}\xi^2\right]\right\}\psi(x)=0, ~~~~\epsilon=2E.
\end{equation}

Take the wave functions of the form
\begin{equation}\label{}
\psi(x)=e^{\frac{\xi\cosh^2(x)}{2}}y(x).
\end{equation} Substituting this into Eq. (\ref{psi}) allows us to obtain
\begin{equation}\label{y(x)}
y''(x)+\xi  \sinh (2 x) y'(x)+[(m+2) \xi  \cosh (2 x)+\epsilon ]y(x)=0.
\end{equation}

Take a new variable $z=\cosh^2(x)$. The above equation becomes
\begin{equation}\label{}
4 (z-1) z y''(z)+[4 z (\xi  (z-1)+1)-2] y'(z)+((m+2) \xi  (2 z-1)+\epsilon )y(z)=0.
\end{equation} which can be re-arranged as
\begin{equation}\label{y-heun}
y''(z)+\left[\xi +\frac{1}{2} \left(\frac{1}{z}+\frac{1}{z-1}\right)\right] y'(z)+\frac{(m+2) \xi  (2 z-1)+\epsilon }{4 (z-1) z}y(z)=0.
\end{equation}
When comparing this with the confluent Heun differential equation in the simplest uniform form [13]
\begin{equation}\label{}
\frac{d^2H(z)}{dz^2}+\left(\alpha+\frac{1+\beta}{z}+\frac{1+\gamma}{z-1}\right)\frac{dH(z)}{dz}+\left(\frac{\mu}{z}+\frac{\nu}{z-1}\right)H(z)=0,
\end{equation}
we find the solution to (\ref{y-heun}) is given by the acceptable confluent Heun function
$H_{c}(\alpha, \beta, \gamma, \delta, \eta; z)$ with
\begin{equation}\label{}
\alpha=\xi, ~~~\beta=-\frac{1}{2}, ~~~ \gamma=-\frac{1}{2}, ~~~\mu=\frac{\xi(m+2)-\varepsilon}{4}, ~~~\nu=\frac{\xi(m+2)+\varepsilon}{4},
\end{equation}from which we are able to calculate the parameters $\delta$ and $\eta$ involved in
$H_{c}(\alpha, \beta, \gamma, \delta, \eta;z)$ as
\begin{equation}\label{}
\delta=\mu+\nu-\frac{1}{2}\alpha(\beta+\gamma+2)=\frac{1}{2} (m+1) \xi, ~~~~
\eta=\frac{1}{2}\alpha(\beta+1)-\mu-\frac{1}{2}(\beta+\gamma+\beta\gamma)=\frac{1}{8} [-2 (m+1) \xi +2 \epsilon +3].
\end{equation}

It is found that the parameter $\eta$ related to energy levels
is involved in the confluent Heun function. The wave function
given by this function seems to be analytical, but the key issue is
how to first get the energy levels. Otherwise, the solution
becomes unsolvable. Generally, the confluent Heun function can be expressed as a series expansions
\begin{equation}\label{}
H_C(\alpha, \beta, \gamma, \delta, \eta, z)=\sum\limits_{n=0}^{\infty}v_n(\alpha, \beta, \gamma, \delta,
\eta, \xi)z^n, ~~~~|z|<1.
\end{equation}
The coefficients $v_n$ are given by a three-term recurrence relation
\begin{equation}\label{}
A_n v_n-B_n v_{n-1}-C_n v_{n-2}=0, ~~~v_{-1}=0, ~~~v_{0}=1,
\end{equation} with
\begin{equation}\label{}
\begin{array}{l}
A_n=1+\displaystyle\frac{\beta}{n}, \\[2mm]
B_n=1+\displaystyle\frac{1}{n}(\beta+\gamma-\alpha-1)+\frac{1}{n^2}\left\{\eta-\frac{1}{2}(\beta+\gamma-\alpha)-\frac{\alpha\beta}{2}
+\frac{\beta\gamma}{2}\right\}, \\[2mm]
C_n=\displaystyle\frac{\alpha}{n^2}\left(\frac{\delta}{\alpha}+\frac{\beta+\gamma}{2}+n-1\right).
\end{array}
\end{equation}

To make the confluent Heun functions reduce to polynomials, two termination
conditions have to be satisfied [13, 14]
\begin{equation}\label{con}
\mu+\nu+N\alpha=0, ~~~~\Delta_{N+1}(\mu)=0,
\end{equation} where
\small
{\begin{equation}\label{con2}
\left|
\begin{array}{ccccccc}
\mu-p_{1} & (1+\beta) &0 &\ldots & 0 &0 &0\\
N\alpha &\mu-p_{2}+\alpha &2(2+\beta)&\ldots &0 &0 &0\\
0 &(N-1)\alpha &\mu-p_{3}+2\alpha &\ldots &0 &0 &0\\
\vdots &\vdots&\vdots&\ddots&\vdots&\vdots&\vdots\\
0&0&0&\ldots&\mu-p_{N-1}+(N-2)\alpha &(N-1)(N-1+\beta)&0\\
0&0&0&\ldots&2\alpha &\mu-p_N+(N-1)\alpha & N(N+\beta)\\
0&0&0&\ldots&0&\alpha &\mu-p_{N+1}+N{\alpha}
\end{array}
\right|=0
\end{equation}
}
with
\begin{equation}\label{}
p_N=(N-1)(N+\beta+\gamma).
\end{equation}

For present problem, it is not difficult to see that the first condition is violated. That is, $\mu+\nu+\alpha=0$ when $N=1$. From this we have $m=-4$. This is contrary to the fact $m$ is a positive integer. Therefore we cannot use this method to obtain the eigenvalues. On the other hand, we know that $z\in [1, \infty)$. Thus, the series expansion method is invalid.
This is unlike previous study [22, 24], in which the quasi-exact wave functions and eigenvalues can be obtained by studying those two constraints. The present case is similar to our previous study [20, 21], in which some constraint is violated. We have to choose other approach to study the eigenvalues as used in [20,21].

\section{Fundamental properties}
In this section we are going to study some basic properties of the
wave functions as shown in Figs. 2-4. We first consider the positive integer $m$. Since the energy spectrum
cannot be given explicitly we have to solve the second order
differential equation (\ref{psi}) numerically. We denote the energy levels
as $\epsilon_{i}(i\in[1,6])$ in Table 1. We find that
the energy levels $\epsilon_{i}$ decrease with the
increasing $m$. Originally, we wanted to calculate the energy
levels numerically by using powerful MAPLE, which includes some
special functions such as the confluent Heun function that cannot be found in MATHEMATICA. As we know, the wave
function is given by $\psi(z)=\exp(z\xi/2)H_{c}(\alpha, \beta, \gamma, \delta, \eta, z)$. Generally speaking, the wave function
requires $\psi(z)\to 0$ when $z\to \infty$, i. e. $x\to \infty$. Unfortunately, the present study is unlike our previous study [20,21], in which
$z\to 1$ when $x$ goes to infinity. The energy spectra can be calculated by series expansions through taking $z\to 1$.
On the other hand, the wave functions have a definite parity, e.g  for $m=0$ some wave functions are symmetric. It is found that such properties are violated
when the potential parameter $m$ becomes larger as shown in Fig. 4. That is, the wave
functions for $m=12$ is nonsymmetric. In addition, on the contrary to the case discussed by Razavy [18], in which he supposed the $m$ is taken as positive integers, we
are going to show what happens to the negative $m$ case. We display the graphics in Figs. 5 and 6 for this case. We find that the wave functions are shrunk towards the origin. This makes the amplitude of the wave function is increasing.

\section{Conclusions}
In this work we have studied the quantum system with the Razavy potential, which is symmetric with respect to the variable $x$,
and shown how its exact solutions are found by transforming the original differential
equation into a confluent type Heun differential equation. It is found that the solutions can be expressed by the confluent Heun functions $Hc(\alpha, \beta, \gamma, \delta, \eta)$, in which
the energy levels are involved inside the parameter $\eta$. This makes us to calculate the eigenvalues numerically.
The properties of the wave functions depending on $m$ are illustrated graphically for a given potential
parameter $\xi$. We have found that the even and odd wave functions with definite parity are changed to odd and even wave functions when the potential parameter $m$ increases.
This arises from the fact that the parity, which is a defined symmetry for very small $m$, is completely violated for large $m$.
We have also noticed that the energy levels $\epsilon_{i}$ decrease with the increasing potential parameter $m$.

\vskip 5mm
\noindent
{\Large \bf Competing Interests}\\
The authors declare that there is no conflict of interests
regarding the publication of this paper.

\vspace{5mm} \noindent {\Large \textbf{Acknowledgments}}: This work is
supported by project 20180677-SIP-IPN, COFAA-IPN, Mexico and partially by the CONACYT project under grant
No. 288856-CB-2016 and partially by NSFC with Grant No. 11465006.

\newpage

\begin{table}
\caption{Energy levels of the Schr\"{o}dinger equation with potential (\ref{pon})}
\begin{center}
{\footnotesize
\begin{tabular}{|l|l|l|l|l|l|l|}
\hline
\hline
$\nu$ & $\epsilon_{1}$ & $\epsilon_{2}$ & $\epsilon_{3}$ & $\epsilon_{4}$ & $\epsilon_{5}$
& $\epsilon_{6}$  \\

\hline
$m=-6$ &21.6608 & 35.7557 & 51.3448 &68.3341 & 86.6500 & 106.233  \\
\hline
$m=-5$ &18.1891 & 31.3844 & 46.1503 &62.3746 & 79.9715 & 98.8740  \\
\hline
$m=-4$ &14.6806 & 26.9167 & 40.8214 & 56.2549 & 73.1150 & 91.3249 \\
\hline
$m=-3$ & 11.1259 & 22.3314 & 35.3346 & 49.9525 &66.0599 & 83.5680 \\
\hline
$m=-2$ &7.51110 &17.5996 & 29.6610 &43.4412 & 58.7838 & 75.5860  \\
\hline
$m=-1$ &3.81463 &12.6800 & 23.7644 &36.6914 & 51.2639 & 67.3635  \\
\hline
$m=0$ & 0.00007 & 7.51170 & 17.6027 & 29.6729 & 43.4799 & 58.8919  \\
\hline
$m=1$ & -3.99968 & 2.00200 & 11.1343 & 22.3606 & 35.4208 & 50.1750 \\
\hline
$m=2$ & -8.32288 & -3.99300 & 4.34771 & 14.7494 & 27.0959 & 41.2385 \\
\hline
$m=3$ &-13.2815 &-10.6927&-2.64788 &6.87526 & 18.5501 & 32.1389  \\

\hline
$m=4$ &-19.5196 &-9.46859&-1.17161 &9.87916 & 22.9677 & 38.0537  \\

\hline
$m=5$ &-27.7547 &-15.7094&-9.29612 &1.24110 & 13.8439 & 28.5940  \\

\hline
$m=6$ &-38.0314 &-21.6913&-17.5131 &-7.12621 & 4.89289 & 19.3065  \\

\hline
$m=7$ &-49.9928 &-28.2027&-25.9897 &-14.8827 & -3.78434 & 10.2625  \\

\hline
$m=8$ &-63.3335 &-35.8866&-21.7455 &-12.1464 & 1.51447 & 17.5661  \\

\hline
$m=9$ &-77.8339 &-44.5255&-27.8571 &-20.2355 & -6.89162 & 8.76577  \\

\hline
$m=10$ &-93.3024 &-54.9017&-33.6970 &-28.1690 & -14.8944 & 0.229704  \\

\hline
$m=11$ &-109.592 &-65.743&-39.7373 &-36.1005 & -22.4007 & -8.04337  \\

\hline
$m=12$ &-126.580 &-77.2416&-46.3335 &-29.3139 & -16.0647 & 1.06475  \\

\hline\hline
\end{tabular}
}
\end{center}
\end{table}

\begin{figure}[ht]
\includegraphics[width=10cm]{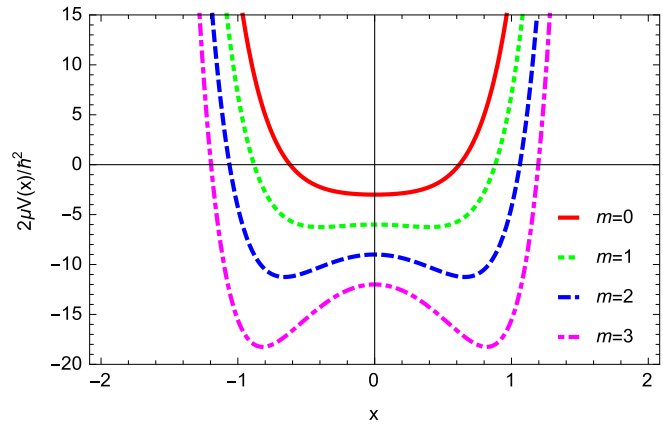}\hfil%
\caption{\label{f1}(Color online) A plot of potential  as function of the variables $x$ and $m$. }
\end{figure}

\begin{figure}
\centering
\includegraphics[width=16cm]{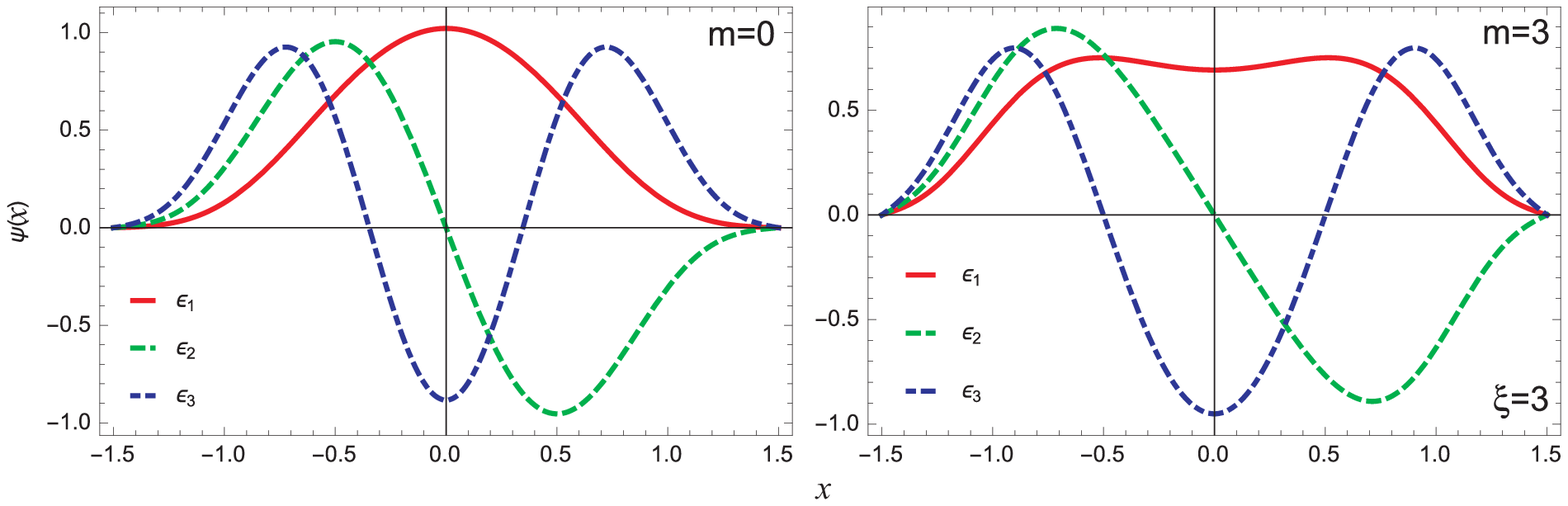}\hfil%
\caption{\label{f1}(Color online) The characteristics of the potential ${\cal V}(z)$ as a function of the position $z$. We take $m=0, 1$ and $\xi=3$. }
\end{figure}

\begin{figure}
\centering
\includegraphics[width =16cm]{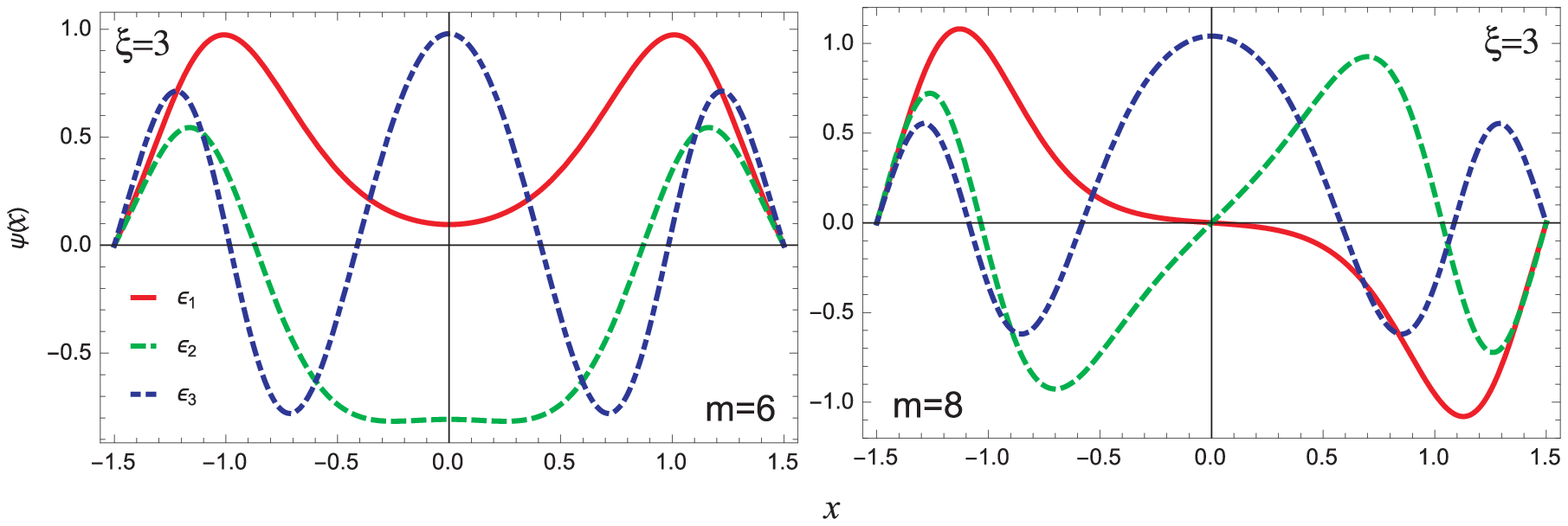}
\caption{(Color online) The characteristics of the potential ${\cal V}(z)$ as a function of the position $z$. We take $m=6, 8$ and $\xi=3$. }
\end{figure}

\begin{figure}
\centering
\includegraphics[width =16cm]{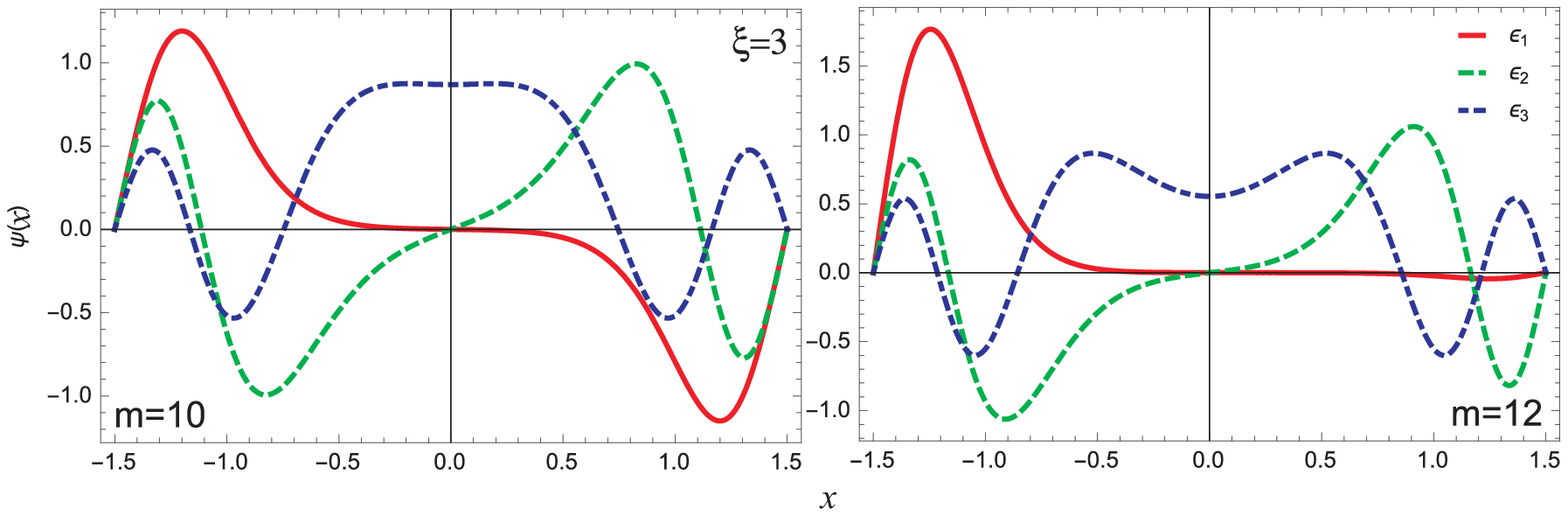} (3)
\caption{(Color online)Same as above case but $m=10, 12$. }
\end{figure}

\begin{figure}
\centering
\includegraphics[width =16cm]{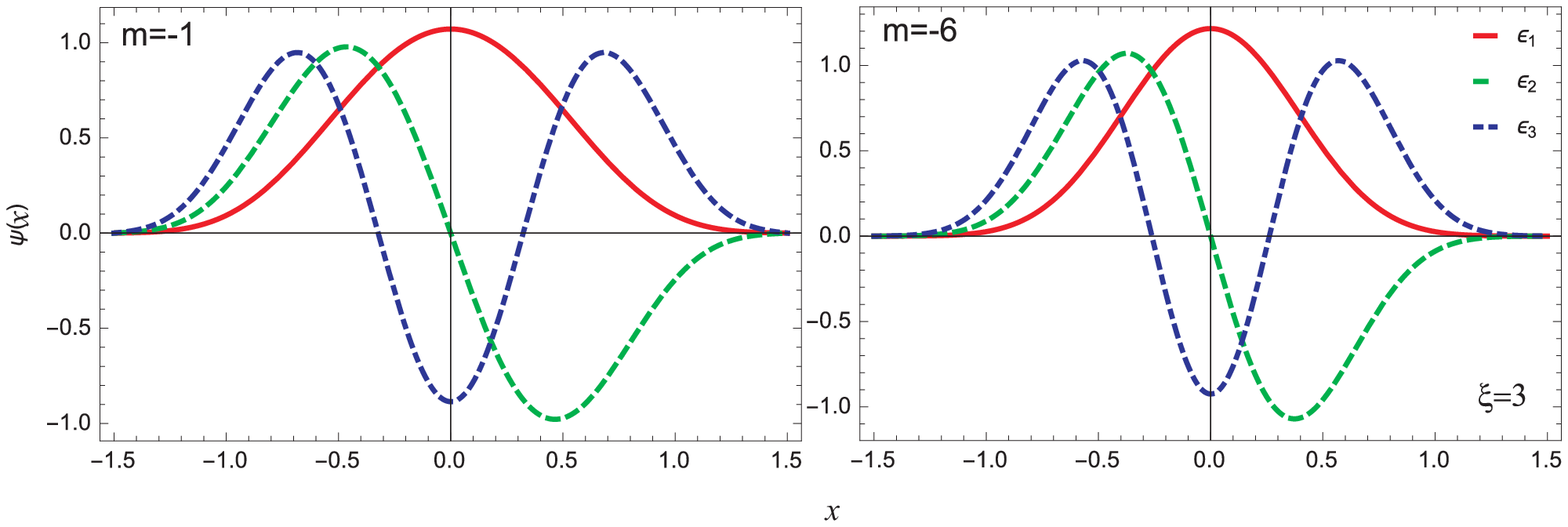}
\caption{(Color online) The characteristics of the potential ${\cal V}(z)$ as a function of the position $z$. We take $m=-1, -6$ and $\xi=3$. }
\end{figure}

\begin{figure}
\centering
\includegraphics[width =16cm]{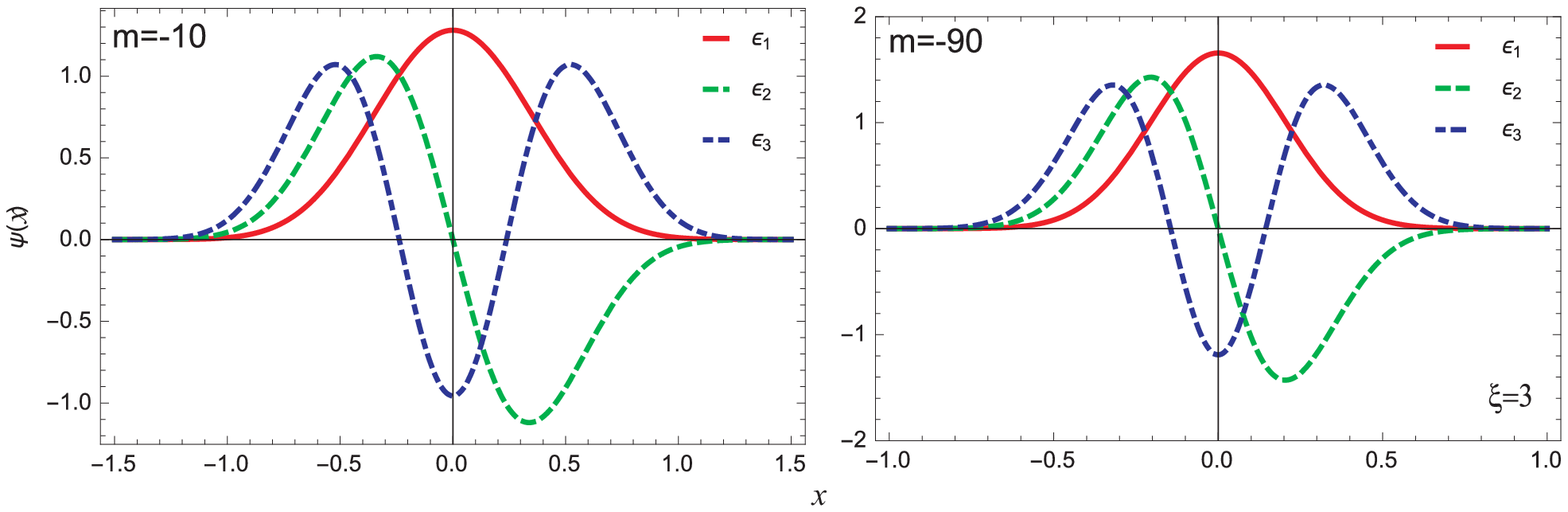}
\caption{(Color online)Same as above case but $m=-10,-90$. }
\end{figure}

\end{document}